\documentclass[preprint,aps,12pt,showpacs,nofootinbib,tightenlines,amsmath,amssymb]{revtex4}

\usepackage{txfonts}
\usepackage{graphicx}
\usepackage{dcolumn}
\usepackage{bm}
\usepackage{amssymb}
\usepackage{latexsym}

\begin{document}

\title{\bf SU(3) Gauge Family Symmetry and Prediction for the Lepton-Flavor Mixing and Neutrino Masses with Maximal Spontaneous CP Violation  }
\author{Yue-Liang~Wu}
\email[Electronic address: ]{ylwu@itp.ac.cn}
\affiliation{State Key Laboratory of Theoretical Physics (SKLTP)\\
Kavli Institute for Theoretical Physics China (KITPC) \\
Institute of Theoretical Physics, Chinese Academy of Sciences, Beijing 100190, China}

\date{\today}

\begin{abstract}
A model for the lepton-flavor mixing and CP violation is proposed based on the SU$_F$(3) gauge family symmetry and the Majorana feature of neutrinos. A consistent prediction for the lepton-flavor mixing and masses is shown to be resulted from the appropriate vacuum structure of SU$_F$(3) gauge symmetry breaking. By choosing the SU$_F$(3) gauge fixing condition to possess a residual $Z_2$ symmetry and requiring the vacuum structure of spontaneous symmetry breaking to have approximate global U(1) family symmetries, we obtain naturally the tri-bimaximal mixing matrix and largely degenerate neutrino masses in the neutrino sector and the small mixing matrix in the charged-lepton sector. With a simple ansatz that all the smallness due to the approximate global U(1) family symmetries is characterized by a single Wolfenstein parameter $\lambda \simeq 0.22$, and the charged-lepton mixing matrix has a similar hierarchy structure as the CKM quark mixing matrix, we arrive at a consistent prediction for the MNSP lepton-flavor mixing with a maximal spontaneous CP violation:  $\delta =\pi/2$, $\sin^2\theta_{13} \simeq \frac{1}{2}\lambda^2 \simeq 0.024$ ($\sin^22\theta_{13} \simeq 0.094$), $\sin^2\theta_{12} \simeq \frac{1}{3}(1 - 2\lambda^3) \simeq 0.326$ and $\sin^2\theta_{23} \simeq \frac{1}{2}(1 - \lambda^2) \simeq 0.48$, which agree well with the current experimental data. The CP-violating Jarlskog-invariant is obtained to be $J_{CP} \simeq \frac{1}{6}\lambda(1-\lambda^2/2-\lambda^3)\sin\delta \simeq 0.035$, which is detectable in next generation neutrino experiment. The small masses of the neutrinos and charged leptons are simply attributed to the standard seesaw mechanism. The largely degenerate neutrino masses with the normal hierarchy and inverse hierarchy are discussed and found be at the order $m_{\nu_i} \simeq O(\lambda^2) \simeq 0.04\sim 0.06$ eV with a total mass $\sum m_{\nu} \sim 0.15$ eV, which is testable in future precision astrophysics and cosmology.
\end{abstract}

\maketitle


The nonzero neutrino masses strongly indicate new physics beyond the standard model(SM). In addition to the Cabbibo-Kobayashi-Maskawa(CKM)\cite{CKM} quark mixing in the SM, there exists also the Maki-Nakagawa-Sakata-Pontecorvo(MNSP)\cite{MNSP} lepton-flavor mixing. The observed neutrino experimental data\cite{SuperK,SNO,KamLAND,Soudan,MARCO,K2K,SK,GNO,CHOOZ,T2K,MINOS} can well be described by neutrino oscillations via three neutrino mixings with massive neutrinos\cite{N1,N2,FIT1,FIT2}. The current neutrino experiments have paid attention to the measurement of the mixing angle $\theta_{13}$ and the improvement on the accuracy of the oscillation parameters. The global fits from various experimental data lead to the following constraints on the mass-squared differences and the three mixing angles given in\cite{FIT1}:
\begin{eqnarray}
& &  \Delta m_{21}^{2}=m_{\nu_2}^2 - m_{\nu_1}^2 = 7.58^{+0.22}_{-0.26}\times {10}^{-5}\ {\mathrm{eV}}^{2}, \quad \Delta m_{32}^{2}= m_{\nu_3}^2 - m_{\nu_1}^2  = 2.35^{+0.12}_{-0.09}\times {10}^{-3}\ {\mathrm{eV}}^{2} \nonumber \\
& &  \sin^2\theta_{12} = 0.312^{+0.017}_{-0.016},\qquad \sin^2\theta_{23} = 0.42^{+0.08}_{-0.03},\qquad  \sin^2\theta_{13}= 0.025\pm 0.007
\end{eqnarray}
and in\cite{FIT2}
\begin{eqnarray}
& &  \Delta m_{21}^{2}=m_{\nu_2}^2 - m_{\nu_1}^2 = 7.59^{+0.20}_{-0.18}\times {10}^{-5}\ {\mathrm{eV}}^{2}, \quad \Delta m_{31}^{2}= m_{\nu_{3}}^2 - m_{\nu_1}^2  = 2.50^{+0.09}_{-0.16}\times {10}^{-3}\ {\mathrm{eV}}^{2} \nonumber \\
& &  \sin^2\theta_{12} = 0.312^{+0.015}_{-0.017},\qquad \sin^2\theta_{23} = 0.52^{+0.06}_{-0.07},\qquad  \sin^2\theta_{13}= 0.013^{+0.007}_{-0.005}
\end{eqnarray}
Both the mixing angles $\theta_{12}$ and $\theta_{23}$ have been determined to be more precise than the mixing angle $\theta_{13}$. Recently, the Daya Bay Reactor Neutrino Experiment has directly measured a non-zero value for the neutrino mixing angle $\theta_{13}$ with a significance of 5.2 standard deviations\cite{An:2012eh}
\begin{equation}
\sin^22\theta_{13}=0.092\pm 0.016({\rm stat})\pm0.005({\rm syst}), \quad (\sin^2\theta_{13}\simeq 0.024\pm 0.004 \pm 0.001)
\end{equation}
which is analyzed in a three-neutrino framework.

A non-zero value for the $\theta_{13}$ plays a fundamental role on the search for CP violation in the lepton sector, which may help us to understand the origin of CP violation via spontaneous symmetry breaking\cite{TDL}. Phenomenologically, such mixing angles slightly deviate from the so-called tri-bimaximal mixing\cite{HPS} with $\theta_{12} = \sin^{-1}(1/\sqrt{3}) = 35^{\circ}$, $\theta_{23} = \sin^{-1}(1/\sqrt{2}) = 45^{\circ}$ and $\theta_{13} =0$. In comparison with the quark masses and CKM quark mixing which has a hierarchy structure characterized by the Wolfenstein parameter $\lambda$\cite{LW}, it raises a puzzle that why neutrino masses are so tiny or largely degenerate, but their mixing angles are so large and even maximal. Thus revealing the origin of large mixing angles and small masses of neutrinos is important not only for understanding neutrino physics, but also for exploring new physics beyond the SM. Great theoretical efforts have been made to study such an intriguing mixing matrix and analyze the possible nonzero $\theta_{13}$ and CP violation via various symmetry and phenomenological considerations\cite{HPS1,HPS2,HPS3,HPS4,TBM1,TBM2,TBM3,TBM4,TBM5,TBM6,TBM7,TBM8,TBM9,TBM10, TBM11,TBM12,TBM13,TBM14,TBM15,TBM16,TBM17, FLee,HWW,Wu:2008zzj,Wu:2008ep,LFM1,LFM2,LFM3,LFM4,LFM5,LFM6,LFM7,LFM8,LFM9,LFM10, LFM11,LFM12,LFM13, LFM14,LFM15, LFM16,LFM17,LFM18,LFM19,LFM20,LFM21,LFM22,LFM23,LFM24,LFM25,LFM26,LFM27,LFM28,LFM29,LFM30,LFM31,LFM32,LFM33,LFM34,LFM35,LFM36,LFM37,
LFM38}. It is interesting to notice that the only peculiar property for neutrinos is that they can be Majorana fermions, so a natural explanation for the puzzle would be attributed to the Majorana features of neutrinos, which strongly motivates us to go beyond the standard model(SM). The greatest success of the SM is the gauge symmetry structure $SU_{c}(3)\times SU_{L}(2)\times U_{Y}(1)$, which has been tested by more and more precise experiments. As a simple extension of the SM with three families and Majorana neutrinos, a non-abelian gauge family symmetry SO(3) has been builded to explore the lepton-flavor mixing for the maximal mixing between muon-neutrino and tau-neutrino as well as the possible nearly bi-maximal and tri-bimaximal neutrino mixings\cite{Wu:2008zzj,Wu:2008ep,YLW1,YLW2,YLW3,YLW4,YLW5,YLW6,CS,MA,CW,BHKR}. In refs.\cite{Wu:2008zzj}, the nearly tri-bimaximal neutrino mixing matrix was generally obtained from diagonalizing a $Z_3$ symmetric mass matrix after SO(3) gauge symmetry breaking and it provided a favorite prediction for the nonzero mixing angle $\theta_{13}$ with $\sin^2\theta_{13} \simeq 0.017$ .

In this note, we are going to extend the SO(3) gauge family symmetry to the SU$_F$(3) gauge family symmetry which was first introduced in early time for estimating the top quark mass\cite{Yanagida:1979gs}. It is different from the consideration in ref.\cite{Wu:2008zzj} where a triplet vector-like heavy Majorana neutrino $N=(N_1,N_2,N_3))$ was introduced by imposing the Majorana condition $N_i=N_i^c$, thus only SO(3) symmetry rather than SU$_F$(3) symmetry was allowed due to the real representation of Majorana neutrinos. Here we will not introduce the triplet vector-like heavy Majorana neutrinos and only consider the right-handed heavy neutrinos with Majorana type Yukawa interactions. As a consequence, we will show how the SU$_F$(3) gauge family symmetry enables us to construct a simple gauge family model for understanding the lepton-flavor mixing and masses. For the purpose in our present consideration, the SU$_F$(3) gauge family model contains only a minimal set of new particles beyond the SM, which includes the SU$_F$(3) gauge bosons, the right-handed SU$_F$(3) triplet neutrino field $N_R$ and a vector-like SU$_F$(3) triplet charged-lepton field $E$, two SU$_F$(3) tri-triplet Higgs bosons $\Phi_{\nu}$ and $\Phi$, and a real singlet Higgs boson $\phi_s$.

With the above mentioned minimal set of new fields, we get the following $SU_F(3)\times SU(2)_{L}\times U(1)_{Y}$ invariant Yukawa interactions for the neutrinos and charged-leptons,
\begin{eqnarray}
{\cal L}_Y & = &  y_L^{\nu} \bar{l} \tilde{H} N_R  + \frac{1}{2}\xi^{\nu} \bar{N}_R \Phi_{\nu} N_R^c   + y_L^e \bar{l} H E + y_R^e \bar{e}_R  \phi_s E  +  \frac{1}{2}\xi^e \bar{E} \Phi E + H.c. \label{LY0}
\end{eqnarray}
where $y_L^{\nu}$, $y_L^e$, $y_R^e$, $\xi^e$ and $\xi^{\nu}$ are the
real Yukawa coupling constants. All the fermions $\nu_{Li}$, $e_{Li}$, $e_{Ri}$, $N_{Ri}$ and $E_i$ $(i=1,2,3)$ belong to the SU$_F$(3) triplets. $\bar{l}_i= (\bar{\nu}_{Li}, \bar{e}_{Li} )$ denote $SU_L(2)$ doublet leptons, $N_{Ri}$ are $SU_L(2)$ singlet right-handed neutrinos with $N_{Ri}^c = c\bar{N}_{Ri}^T$. $H$ is the $SU_L(2)$ doublet Higgs boson with $\tilde{H} = \tau_2 H^*$.  $E_i$ are $SU_L(2)$ singlet vector-like charged leptons. The scalar fields $\Phi_{\nu}$ and $\Phi$ are two SU$_F$(3) tri-triplets Higgs bosons. The hermiticity condition of the above Lagrangian and the Majorana feature of the right-handed neutrinos imply that
\begin{eqnarray}
\Phi_{\nu} = \Phi_{\nu}^T, \quad \Phi = \Phi^{\dagger}\ .
\end{eqnarray}
Namely $\Phi_{\nu}$ is a complex symmetric tri-triplet Higgs boson and $\Phi$ is an Hermitian
tri-triplet Higgs boson. They transform under the SU$_F$(3) gauge transformation $g(x) \in SU_F(3)$ as follows
\begin{eqnarray}
\Phi_{\nu} \to  g\Phi_{\nu}g^T, \quad \Phi \to g \Phi g^{\dagger}\ .
\end{eqnarray}
Thus the Lagrangian in Eq.(\ref{LY0}) is the most general one ensured by the $SU_F(3)\times SU(2)_{L}\times U(1)_{Y}$ gauge symmetry with an additional $Z_2$-parity symmetry: $\phi_s \to - \phi_s$, $e_R \to - e_R$. We will show that the smallness of neutrino masses and nearly tri-bimaximal neutrino mixing can well be understood simultaneously via an appropriate vacuum structure of the SU$_F$(3) tri-triplet Higgs bosons, which is yielded by just requiring a residual $Z_2$ symmetry and approximate global U(1) family symmetries\cite{HW,Wu:1994ja,Wolfenstein:1994jw,Wu:1994vx}.

Before proceeding, let us first discuss the features of SU$_F$(3) gauge symmetry. In
terms of the SU$_F$(3) representation, one can reexpress the complex symmetric
tri-triplet Higgs boson $\Phi_{\nu}$ into the following general form
\begin{eqnarray}
& & \Phi_{\nu}  \equiv  U_{\nu} \phi_{\nu} U_{\nu}^T, \quad
U_{\nu}(x) = e^{i\lambda^a \Theta^{\nu}_a(x)} \label{GFC}
\end{eqnarray}
with $\lambda^a$ $(a=1,\cdots, 8)$ being the generators of SU$_F$(3). Where
$\Theta_a^{\nu}(x)$ $(i=1,\cdots,8)$ will correspond to the eight scalar fields of SU$_F$(3),
and $\phi_{\nu}$ is a real symmetric matrix consisting of three scalar fields $\phi_i^{\nu}(x)$ $(i=1,2,3)$.
There are in general two types of non-trivial structures for $\phi_{\nu}(x)$. One is
\begin{eqnarray*}
\phi_{\nu}(x) =  \phi^{Z_3}_{\nu}(x)
= \left(
                     \begin{array}{ccc}
                       \phi_1^{\nu} & \phi_2^{\nu} & \phi_3^{\nu} \\
                       \phi_2^{\nu} & \phi_3^{\nu} & \phi_1^{\nu} \\
                       \phi_3^{\nu} & \phi_1^{\nu} & \phi_2^{\nu} \\
                     \end{array}
                   \right) \ ,
\end{eqnarray*}
which is in a cyclic permuted form $[\phi_{\nu}(x)]_{ij} =
\phi^{\nu}_{i+j-1}(x)$ with $(i+j-1)$ mod.3, namely it has a $Z_3$ symmetry for the cyclic permutation among the three components $(\phi^{\nu}_1,\phi^{\nu}_2,\phi^{\nu}_3) $. And the other is
\begin{eqnarray}
\phi_{\nu}(x) = \phi^{Z_2}_{\nu}(x)
\equiv \phi_0^{\nu}(x) I_3 +\phi^{Z_3Z_2}_{\nu}(x) = \left(
                     \begin{array}{ccc}
                     \phi^{\nu}_0 + \phi_1^{\nu} & \phi_2^{\nu} & \phi_2^{\nu} \\
                       \phi_2^{\nu} &  \phi^{\nu}_0 + \phi_2^{\nu} & \phi_1^{\nu} \\
                       \phi_2^{\nu} & \phi_1^{\nu} &  \phi^{\nu}_0 + \phi_2^{\nu} \\
                     \end{array}
                   \right) \ ,
\end{eqnarray}
which has a $Z_2$ symmetry for the permutation between the matrix elements $[\phi_{\nu}]_{ij} = [\phi_{\nu}]_{ji}$ with $i\ \mbox{and/or}\ j= 2,3$. Where $I_3$ is the $3\times 3$ unit matrix and the field $\phi^{Z_3Z_2}_{\nu}=  \phi^{Z_3}_{\nu} (\phi_1^{\nu}, \phi_2^{\nu}, \phi_3^{\nu}=\phi_2^{\nu}) $ possesses both $Z_2$ symmetry and $Z_3$ symmetry. In general, there are three $Z_2$ symmetric matrices. The other two $Z_2$ symmetric matrices have similar property: $[\phi_{\nu}]_{ij} = [\phi_{\nu}]_{ji}$ with $i\ \mbox{and/or}\ j= 1,2$, and $[\phi_{\nu}]_{ij} = [\phi_{\nu}]_{ji}$ with $i\ \mbox{and/or}\ j= 1,3$.

The above two types of structure may be regarded as the unique property of the cyclic Abelian finite groups $Z_3$ and $Z_2$ for any real symmetric matrix field $\phi_{\nu}(x)$ containing three scalar fields. They are actually the nontrivial invariant subgroups of the non-Abelian symmetric group $S_3=\{ t_i,\ T_i \}$ $(i=1,2,3)$ with
\begin{align*}
t_{1}  &  \equiv T_0 =\left(
\begin{array}
[c]{ccc}%
1 & 0 & 0\\
0 & 1 & 0\\
0 & 0 & 1
\end{array}
\right) ,\,\  t_{2}=\left(
\begin{array}
[c]{ccc}%
0 & 1 & 0\\
0 & 0 & 1\\
1 & 0 & 0
\end{array}
\right) ,\,\  t_{3}=\left(
\begin{array}
[c]{ccc}%
0 & 0 & 1\\
1 & 0 & 0\\
0 & 1 & 0
\end{array}
\right)  ,
\end{align*}
\begin{align*}
T_{1}  &  =\left(
\begin{array}
[c]{ccc}%
1 & 0 & 0\\
0 & 0 & 1\\
0 & 1 & 0
\end{array}
\right),\,\  T_{2}=\left(
\begin{array}
[c]{ccc}%
0 & 1 & 0\\
1 & 0 & 0\\
0 & 0 & 1
\end{array}
\right),\, \   T_{3}  =\left(
\begin{array}
[c]{ccc}%
0 & 0 & 1\\
0 & 1 & 0\\
1 & 0 & 0
\end{array}
\right) \ ,
\end{align*}
which give the explicit three dimensional unitary representations for $Z_3$ subgroup with $Z_3=\{ t_i \}$ $(i=1,2,3)$ and three $Z_2$ subgroups with $Z_2 = \{T_0, T_1\}, \{T_0, T_2\}, \{T_0, T_3\}$. A large number of papers have adopted the $S_3$ symmetry to construct some interesting  models\cite{S3M1,S3M2,S3M3,S3M4,S3M5,S3M6,S3M7,S3M8,S3M9,S3M11,S3M12,S3M13,S3M14,S3M15,S3M16,S3M17,S3M18,S3M19,S3M20,
S3M21,S3M22,S3M23,S3M24,S3M25}.

It can easily be checked that the non-trivial structures of
$\phi_{\nu}(x)$ with three scalar fields can explicitly be expressed in terms of the group
representation $\{ T_i |i=0,1,2,3\}$ as follows
\begin{eqnarray*}
\phi^{Z_3}_{\nu}(x) = \phi_1^{\nu}(x) T_1 +   \phi_2^{\nu}(x) T_2 +
\phi_3^{\nu}(x) T_3
\end{eqnarray*}
which is invariant under the $Z_3$ operation
\[t_i \phi^{Z_3}_{\nu}(x) t_i = \phi^{Z_3}_{\nu}(x) \]
and
\begin{eqnarray}
\phi^{Z_2}_{\nu}(x) = \phi_0^{\nu}(x) T_0 + \phi_1^{\nu}(x) T_1  +   \phi_2^{\nu}(x) (T_2 + T_3)
\end{eqnarray}
which is invariant under the $Z_2$ operation
\begin{eqnarray}
T_1 \phi^{Z_2}_{\nu}(x) T_1  =  \phi^{Z_2}_{\nu}(x)
\end{eqnarray}
where
\begin{eqnarray}
\phi^{Z_3Z_2}_{\nu}(x) = \phi_1^{\nu}(x) T_1  +   \phi_2^{\nu}(x) (T_2 + T_3)
\end{eqnarray}
is invariant under both the $Z_3$ and $Z_2$ operations
\begin{eqnarray}
t_i \phi^{Z_3Z_2}_{\nu}(x)t_i  =  \phi^{Z_3Z_2}_{\nu}(x), \qquad T_1 \phi^{Z_3Z_2}_{\nu}(x) T_1  =  \phi^{Z_3Z_2}_{\nu}(x)
\end{eqnarray}

We will see below that only the $Z_2$ symmetric matrix field $[\phi_{\nu}]_{ij} = [\phi_{\nu}]_{ji}$ with $i \ \mbox{and/or}\ j= 2,3$ is reliable to explain the current observed neutrino masses and mixing. Therefore, in the following discussions, we will focus on the non-trivial structure of $\phi_{\nu}(x)$ with the $Z_2$ symmetry $\phi_{\nu}(x) = \phi^{Z_2}_{\nu}(x)$.

The SU$_F$(3) gauge invariance allows us to choose an appropriate gauge fixing condition, by making SU$_F$(3)
gauge transformation $g(x)$ to satisfy the condition $g(x) \equiv U_{\nu}(x) \in SU_F(3)$ with $U_{\nu}(x)$ defined in Eq. (\ref{GFC}), we arrive at the following Yukawa interactions with a special gauge fixing condition
\begin{eqnarray}
{\cal L}_Y & = &  y_L^{\nu} \bar{l} \tilde{H} N_R  + \frac{1}{2}\xi^{\nu} \bar{N}_R \phi_{\nu} N_R^c   + y_L^e \bar{l} H E + y_R^e \bar{e}_R  \phi_s E  +  \frac{1}{2}\xi^e \bar{E} \hat{\Phi} E + H.c. \label{LY}
\end{eqnarray}
where $\hat{\Phi} = U_{\nu}^{\dagger} \Phi  U_{\nu}$ remains Hermitian
and contains nine independent scalar fields, which can generally be reexpressed in terms of SU$_F$(3) representation as follows
\begin{eqnarray}
 \hat{\Phi} \equiv U_e \phi U_e^{\dagger},
 \quad U_e(x) \equiv P_e O_{e}, \quad \quad O_e(x) =
e^{i\lambda^i \chi^{e}_i(x)},
\end{eqnarray}
with
\begin{eqnarray}
& & P_e(x)  =  \left(
 \begin{array}{ccc}
  e^{i\eta_1^e(x)} & 0 & 0 \\
    0 & e^{i\eta_2^e(x)} & 0 \\
    0 & 0 & e^{i\eta_3^e(x)}
  \end{array}
 \right),\quad \phi(x) = \left(
                     \begin{array}{ccc}
                       \phi_1(x) & 0 & 0 \\
                       0 & \phi_2(x) & 0 \\
                       0 & 0 & \phi_3(x) \\
                     \end{array}
                    \right)
\end{eqnarray}
where $\chi_i^e(x)$ $(i=1,2,3)$ represent three rotational scalar fields with $\lambda^i$ $(i=1,2,3)$ the generators of SO(3) group, $\eta_i^e(x)$ $(i=1,2,3)$ denote three phase scalar fields and $\phi_i(x)$ $(i=1,2,3)$ are the remaining three independent scalar fields.

When all the scalar fields get their vacuum expectation values(VEVs), both SU$_F$(3) and SU$_L(2)$ gauge symmetries and the discrete symmetries are broken down spontaneously. In the above gauge fixing basis, we shall take the triplet Higgs boson $\phi_{\nu}(x)$ to be the $Z_2$ symmetric one $\phi_{\nu}(x) = \phi^{Z_2}_{\nu}(x)$. This is because for the $Z_3$ symmetric one $\phi_{\nu}(x) = \phi^{Z_3}_{\nu}(x)$, when all the field components $\phi^{\nu}_i$ $(i=1,2,3)$ obtain nonzero VEVs, it is easy to check that the resulting neutrino masses cannot explain the observed neutrino oscillations as two neutrino masses become completely degenerate.

With the above analysis, let us consider the following general vacuum structure of scalar fields
\begin{eqnarray}
& & <\phi_i^{\nu}(x)> = V_i^{\nu}\,\ (i=0,1,2), \quad <H(x)> = v, \nonumber \\
& &  <\phi_i(x)> = V_i \, \ \, \ (i=1,2,3), \quad <\phi_s(x)> = v_s, \\
& & <\chi_i^e(x)> = \theta_i^e , \quad <\eta_i^e(x)> = \delta_i^e,\, \ (i=1,2,3). \nonumber
\end{eqnarray}
namely
\begin{eqnarray}
& & <P_e> \equiv P_{\delta}  = diag.( e^{i\delta_1^e},
e^{i\delta_2^e}, e^{i\delta_3^e} ), \quad <O_e> = e^{i\lambda^i
\theta_i^e}
\end{eqnarray}
where $\delta_i^e$ (i=1,2,3) correspond to the three CP-violating phases and $\theta_i^e$ the three rotational angles, they all arise from the VEVs of spontaneous symmetry breaking.

It is interesting to note from the above vacuum structure that if taking the VEVs of the Higgs tri-triplet $\Phi_{\nu}(x)$ to be $V_0^{\nu} > V_i^{\nu}$ $(i=1,2)$, the SU$_F$(3) gauge symmetry will first be broken down to its subgroup SO(3), and then the nonzero VEVs $V_i^{\nu}$ $(i=1,2)$ further break the SO(3) gauge symmetry down to a vacuum structure with the discrete symmetry $Z_2$. In this case, all the SU$_F$(3) gauge bosons and the three right-handed Majorana neutrinos $N_{Ri}$ as well as the three scalar fields $\phi_i^{\nu}(x)$ become massive. When the Higgs doublet $H(x)$ gets VEV $<H(x)> = v$, the SU$_L(2)$ gauge symmetry is broken down and the three left-handed neutrinos in the SM gets masses. When the SU$_F$(3) Higgs tri-triplet $\Phi(x)$ evaluates the VEVs $<\phi_i(x)> = V_i$, the discrete $Z_2$ symmetry is in general broken down in the charged-lepton sector, the three vector-like charged leptons $E_i$ become massive and meanwhile the SU$_F$(3) gauge bosons further receive contributions for their masses. Meanwhile the nonzero VEVs  $<\eta_i^e(x)> = \delta_i^e$ lead to spontaneous CP violation\cite{TDL}. Once the singlet scalar $\phi_s$ gets VEV $<\phi_s(x)> = v_s$, the $Z_2$-parity symmetry is broken down and the three charged leptons in the SM obtain their masses.

When the VEV $V_0^{\nu}$ is much larger than the VEV $v$, namely the right-handed Majorana neutrinos become very heavy, the resulting Majorana-type Yukawa interactions for the left-handed leptons will decouple from the theory. This may be seen from the following explicit effective Yukawa interactions mediated via the heavy Majorana neutrinos
\begin{eqnarray}
\frac{(y^{\nu}_L)^2}{M_N} \bar{l}\tilde{H} \tilde{H}^{T} l^c \to 0 \qquad \mbox{for}
\quad M_N \to \infty
\end{eqnarray}
which indicates that if requiring the VEVs of the Higgs tri-triplets $\Phi_{\nu}(x)$ and $\Phi(x)$ to satisfy the following conditions
\begin{equation}
 |V_i^{\nu}|\gg v \,  (i=0,1,2), \qquad  \theta_i^e \ll 1\, (i=1,2,3)\, ,
 \label{U1}
\end{equation}
the resulting effective Yukawa interactions possess approximate global U(1) family symmetries:
\begin{eqnarray}
& & l_i \to e^{i\alpha_i} l_i,\qquad  e_{Ri} \to e^{i\alpha_i} e_{Ri}, \nonumber \\
& & E_i \to e^{i\alpha_i} E_i,\qquad N_{Ri} \to e^{i\alpha_i} N_{Ri}
\end{eqnarray}
with $\alpha_i$ (i=1,2,3) the U(1) charges. It shows that if applying the mechanism of approximate global U(1) family symmetries\cite{HW,Wu:1994ja,Wolfenstein:1994jw,Wu:1994vx} to the Yukawa interactions after the gauge and discrete symmetries are broken down spontaneously, we are led naturally the standard seesaw mechanism\cite{SS} to explain the smallness of the left-handed neutrino masses, and meanwhile we are also able to understand naturally the smallness of the charged lepton mixing.

With the above vacuum structure, the mass matrices of the neutrinos and charged leptons are given via the following standard seesaw mechanism generated by the right-handed heavy Majorana neutrinos with $|V_i^{\nu}| \gg v$  $(i=0,1,2)$ and the generalized see-saw mechanism due to the heavy vector-like charged leptons $|V_i| \gg v_s$ $(i=1,2,3)$
\begin{eqnarray}
& & M_{\nu} =  m^D_{N} M_N^{-1} m^D_{N}, \qquad M_e =  V_e m_E^D M_E^{-1} m_E^D V_e^{\dagger},
\end{eqnarray}
with
\begin{eqnarray}
& & m^D_{N} = y_L^{\nu} v, \quad m_E^D = \sqrt{y_L^{e} y_R^e v  v_s },\quad V_e = <U_{e}> = P_{\delta} e^{i \lambda^i \theta_i^e } ,
\end{eqnarray}
and
\begin{eqnarray}
 & & M_N = \xi^{\nu} \left(
        \begin{array}{ccc}
          V_0^{\nu} + V_1^{\nu} &  V_2^{\nu} &  V_2^{\nu} \\
           V_2^{\nu} & V_0^{\nu} + V_2^{\nu} &  V_1^{\nu} \\
           V_2^{\nu} &  V_1^{\nu} &  V_0^{\nu} + V_2^{\nu} \\
        \end{array}
      \right),\qquad  M_E = \xi^e \left(
            \begin{array}{ccc}
               V_1 & 0 & 0 \\
              0 &  V_2 & 0 \\
              0 & 0 &  V_3 \\
            \end{array}
          \right) \\
& & V_e^{\dagger} \equiv
              \left(
                     \begin{array}{ccc}
                       c_{12}^ec_{13}^e\ \ & s_{12}^ec_{13}^e\ \ & s_{13}^e \\
                       -s_{12}^ec_{23}^e-c_{12}^es_{23}^es_{13}^e\  \ &
                       c_{12}^ec_{23}^e - s_{12}^es_{23}^es_{13}^e\ \ & s_{23}^e c_{13}^e \\
                       s_{12}^e s_{23}^e - c_{12}^ec_{23}^es_{13}^e\ \ & -c_{12}^es_{23}^e - s_{12}^ec_{23}^es_{13}^e\  \
                       & c_{23}^ec_{13}^e \\
                     \end{array}
                   \right)  P_{\delta}^{\ast},
\end{eqnarray}
where we have used the notations $c_{ij}^e \equiv \cos\theta_{ij}^e$
and $s_{ij}^e \equiv \sin\theta_{ij}^e$. Note that $\theta_{ij}^e$ $(i<j)$
are generally given as the functions of $\theta_i^e$ $(i=1,2,3)$.

Considering the SU$_F$(3) gauge symmetry breaking scenario is via SU$_F$(3) to SO(3) which is realized by requiring the following hierarchy structure
\begin{equation}
 V_0^{\nu} \gg V_1^{\nu},\,  V_2^{\nu},
\end{equation}
which indicates that both the right-handed and left-handed Majorana neutrinos become largely degenerate and the SU$_F$(3) gauge bosons in the coset SU$_F$(3)/SO(3) become heavier than the SO(3) gauge bosons. When further requiring an approximate $Z_2$ symmetry in the charged lepton sector, the VEVs of the Higgs tri-triplet $\Phi(x)$ will have the following hierarchy structure
\begin{equation}
 V_1 \gg V_2,\,  V_3 \, ,
 \label{Z2}
\end{equation}
which will provide a natural explanation for the smallest electron mass via the generalized see-saw mechanism.

To be explicit,  diagonalizing the mass matrices of the neutrinos and charged leptons as follows
\begin{eqnarray}
V_{\nu}^{T} M_{\nu} V_{\nu} = diag.(m_{\nu_e}, m_{\nu_{\mu}},
m_{\nu_{\tau}} ), \quad V_e^{\dagger} M_e V_e = diag.(m_e, m_{\mu},
m_{\tau} ) \, ,
\end{eqnarray}
we obtain naturally a parameterless neutrino mixing matrix
\begin{eqnarray}
 V_{\nu} = \left(
  \begin{array}{ccc}
    \frac{2}{\sqrt{6}}
     & \frac{1}{\sqrt{3}} & 0 \\
    -\frac{1}{\sqrt{6}} & \frac{1}{\sqrt{3}} & \frac{1}{\sqrt{2}} \\
    -\frac{1}{\sqrt{6}} & \frac{1}{\sqrt{3}} & -\frac{1}{\sqrt{2}} \\
  \end{array}
\right),
\end{eqnarray}
which is the so-called tri-bimaximal mixing matrix\cite{HPS}. As a consequence, we arrive at, in the mass
eigenstates, the following MNSP lepton-flavor mixing matrix
\begin{eqnarray}
V_{MNSP} & = & V_e^{\dagger} V_{\nu}  =
                        \left(
                     \begin{array}{ccc}
                       c_{12}^ec_{13}^e\ \ & s_{12}^ec_{13}^e\ \ & s_{13}^e \\
                       -s_{12}^ec_{23}^e-c_{12}^es_{23}^es_{13}^e\  \ &
                       c_{12}^ec_{23}^e - s_{12}^es_{23}^es_{13}^e\ \ & s_{23}^e c_{13}^e \\
                       s_{12}^e s_{23}^e - c_{12}^ec_{23}^es_{13}^e\ \ & -c_{12}^es_{23}^e - s_{12}^ec_{23}^es_{13}^e\  \
                       & c_{23}^ec_{13}^e \\
                     \end{array}
                   \right)    \nonumber \\
 & & \qquad  \times \left(
             \begin{array}{ccc}
               e^{-i\delta_1^e} & 0 & 0 \\
               0 &  e^{-i\delta_2^e}  & 0 \\
               0 & 0 &  e^{-i\delta_3^e}  \\
             \end{array}
           \right)
            \left(
  \begin{array}{ccc}
    \frac{2}{\sqrt{6}}
     & \frac{1}{\sqrt{3}} & 0 \\
    -\frac{1}{\sqrt{6}} & \frac{1}{\sqrt{3}} & \frac{1}{\sqrt{2}} \\
    -\frac{1}{\sqrt{6}} & \frac{1}{\sqrt{3}} & -\frac{1}{\sqrt{2}} \\
  \end{array}
\right) \\
      & \equiv &   P_{\beta}^e \left(
                     \begin{array}{ccc}
                       c_{12}c_{13}\ \ & s_{12}c_{13}\ \ & s_{13}e^{-i\delta} \\
                       -s_{12}c_{23}-c_{12}s_{23}s_{13}e^{i\delta}\  \ &
                       c_{12}c_{23} - s_{12}s_{23}s_{13}e^{i\delta} \ \ & s_{23} c_{13} \\
                       s_{12} s_{23} - c_{12}c_{23}s_{13}e^{i\delta} \ \
                       & -c_{12}s_{23} - s_{12}c_{23}s_{13} e^{i\delta}\  \
                       & c_{23}c_{13} \\
                     \end{array}
                   \right) \left(
             \begin{array}{ccc}
                 e^{i\alpha_1} & 0 & 0 \\
               0 &  e^{i\alpha_2} & 0 \\
               0 & 0 & 1 \\
             \end{array}
           \right) \label{MNSP} \, ,
\end{eqnarray}
where we have used the notations $P_{\beta}^e = diag.\left( e^{i\beta_1^e}, e^{i\beta_2^e},\
e^{i\beta_3^e} \right)$, and $c_{ij}\equiv \cos\theta_{ij}$ and
$s_{ij}\equiv \sin\theta_{ij}$. The phases $\beta_i^e$ (i=1,2,3) and $\alpha_i$ $(i=1,2)$ are
introduced to parameterize the leptonic mixing matrix
into a standard form which has been used widely for CKM quark mixing
matrix. The two phases $\alpha_1$ and $\alpha_2$ are also known as the so-called Majorana
phases for the Majorana neutrinos. Note that the three mixing angles $\theta_{ij}$ and six CP
phases $\beta_i^e$, $\delta$, $\alpha_1$ and $\alpha_2$ are all given by the initial three mixing
angles $\theta_{ij}^e$ $(i<j)$ and three CP phases $\delta_{i}^e$. Formally, the three
phases $\beta_i^e$ can be absorbed by the phase redefinitions of charged leptons, while unlike in the SM, the phases $\beta_i^e$ cannot be rotated away in the model due to the SU$_F$(3) gauge family interactions, their
physical effects will occur in processes involving SU$_F$(3) gauge interactions.

With the smallness of $\theta_{ij}^e$ given in Eq.(\ref{U1}) due to the approximate global U(1) family symmetries after spontaneous symmetry breaking, it is expected that the charged-lepton mixing is similar to the CKM quark mixing and has the hierarchy structure $s_{12}^e \gg s_{23}^e \gg s_{13}^e$. This is because the tri-triplet Higgs $\Phi(x)$ will also have Yukawa interactions in the quark sector which is going to be investigated elsewhere. Thus it is useful to apply the Wolfenstein parametrization\cite{LW} for the charged-lepton mixing matrix,
\begin{eqnarray}
V_e^{\dagger} \simeq \left(
\begin{array}{ccc}
1 - \frac{1}{2} \lambda_e^2 - \frac{1}{8} \lambda_e^4 & \lambda_e & A_e\lambda_e^3 \rho_e  \cr
-\lambda_e  + A_e^2 \lambda_e^5 \left(\frac{1}{2} - \rho_e\right) & 1 - \frac{1}{2} \lambda_e^2 -
\frac{1}{8} \left(1 + 4A_e^2 \right) \lambda_e^4 & A_e\lambda_e^2 \cr
A_e\lambda_e^3 \left(1 - \rho_e\right)  + \frac{1}{2} A_e \lambda_e^5\rho_e
& -A_e\lambda_e^2  + A_e\lambda_e^4 \left( \frac{1}{2} - \rho_e\right)
& 1 - \frac{1}{2} A_e^2 \lambda_e^4 \cr
\end{array}\right)P_{\delta}^{\ast}
\end{eqnarray}
Keeping to the order $O(\lambda_e^3)$, we can reexpress the MNSP lepton-flavor mixing matrix by the following simplified form
\begin{eqnarray}
V_{MNSP} & \simeq & P_{\delta}^{\ast}  \left(
                       \begin{array}{ccc}
                       1 - \frac{1}{2} \lambda_e^2 \ & \lambda_e e^{-i\delta_{e}} \ &  A_e\lambda_e^3 \rho_e e^{-i\delta'_{e}} \\
                       -\lambda_e  e^{i\delta_{e}} \ &  1 - \frac{1}{2} \lambda_e^2  \ &  A_e\lambda_e^2  e^{i(\delta_{e}-\delta'_{e}) } \\
                       A_e\lambda_e^3 \left(1 - \rho_e\right)  e^{i\delta'_{e}} \ &  -A_e\lambda_e^2 e^{-i(\delta_{e}-\delta'_{e}) }  \ & 1 \\
                     \end{array}
                   \right)
           \left(
  \begin{array}{ccc}
    \frac{2}{\sqrt{6}}
     & \frac{1}{\sqrt{3}} & 0 \\
    -\frac{1}{\sqrt{6}} & \frac{1}{\sqrt{3}} & \frac{1}{\sqrt{2}} \\
    -\frac{1}{\sqrt{6}} & \frac{1}{\sqrt{3}} & -\frac{1}{\sqrt{2}} \\
  \end{array}
\right)
\end{eqnarray}
where we have defined the relative phase to be $\delta_{e} = \delta_2^e -\delta_1^e$ and $\delta'_{e} = \delta_3^e -\delta_1^e$.

From the above analysis, we have, in a good approximation, the following relations for the lepton-flavor mixing angles $\theta_{ij}$ and CP-violating phase $\delta$  defined in the standard representation Eq.(\ref{MNSP})
\begin{eqnarray}
& & \sin\theta_{13} \simeq \frac{1}{\sqrt{2}}  \lambda_e |1-A_e\lambda_e^2\rho_e e^{i(\delta_{e}-\delta'_{e})}| , \quad \delta \simeq \delta_{e} \\
& & \sin\theta_{12} \simeq \frac{1}{\sqrt{3}} |1 - \lambda_e^2/2+ \lambda_e e^{-i\delta_{e}} + A_e\lambda_e^3 \rho_e e^{-i\delta'_{e}}|, \\
& & \sin\theta_{23} \simeq \frac{1}{\sqrt{2}} |1- \lambda_e^2/2 - A_e \lambda_e^2 e^{i(\delta_{e}-\delta'_{e})} |
\end{eqnarray}

Note that the approximate global U(1) family symmetries considered in our present case only ensure that the off-diagonal mass matrix elements must be much smaller than the diagonal ones, they do not in general lead to the possible relations between the mixing angles and the mass ratios of quarks and leptons, as those relations require to construct carefully some texture zero mass matrixes. While the global U(1) family symmetries may indicate a hierarchy structure for the mass matrix, for instance, by requiring that the tri-triplet Higgs components $\hat{\Phi}_{ij}$ with large U(1) charges may get small VEVs.

In order to have a quantitative prediction, we make a simple ansatz that the smallness of the charged-lepton mixing due to the approximate global U(1) family symmetries is characterized by a single Wolfenstein parameter $\lambda\simeq 0.22$, and the spontaneous CP violation from the vacuum is maximal, i.e.,
\begin{eqnarray}
\lambda_e\simeq \lambda \simeq 0.22,\quad A_e\simeq 1, \quad \rho_e \simeq 1, \quad \delta \simeq \delta_{e}= \pi/2\, ,\quad \delta'_{e} = \pi
\end{eqnarray}
which leads to the following predictions for the mixing angles $\theta_{ij}$
\begin{eqnarray}
& & \sin^2\theta_{13}\simeq \frac{1}{2} \lambda^2 (1 + \lambda^4 ) \simeq 0.024 \quad (\sin^22\theta_{13} \simeq 0.094 ) , \\
& & \sin^2\theta_{12} \simeq \frac{1}{3} (1- 2\lambda^3+\lambda^4/4) \simeq 0.326,  \\
& & \sin^2\theta_{23} \simeq \frac{1}{2} [ (1 -\lambda^2/2)^2 + \lambda^4 ] \simeq 0.48  ,
\end{eqnarray}
which are consistent with the current experimental data. In particular, the resulting $\theta_{13}$ agrees remarkably with the most recent measurement by the Daya Bay reactor neutrino experiment\cite{An:2012eh}. It strongly implies that the smallness of the charged-lepton mixing matrix is related to the CKM quark mixing matrix, and a maximal spontaneous CP violation is favorite in the lepton sector. Actually, a similar ansatz with $V_e \simeq V_{CKM}$ has been given in ref.\cite{PR}. And also a speculation on $V_{MNSP} \simeq V_{CKM}^{\dagger} U_{TB}$ has been discussed in\cite{AD,KN}.  These models are mainly based on the speculations for the possible deviation of neutrino mixing to the tri-bimaximal mixing ansatz. In our present model we  have provided a more steady theoretical analysis based on some symmetry considerations. 
 
Alternatively, we may take the well determined mixing angle $\theta_{12}$ to extract the CP-violating phase $\delta$. Taking $\sin^2\theta_{12}= 0.312\pm 0.016$, we arrive at the following result
\begin{eqnarray}
& & \delta \simeq (0.55\mp 0.04)\pi, \quad \mbox{or}\quad   \sin\delta \simeq 0.989^{+0.010}_{-0.024}
\end{eqnarray}
which is almost the maximal. As the CP-violating observables should be rephase-invariant, let us define the corresponding Jarlskog-invariant\cite{JCP} $J_{CP}$ in the lepton-flavor mixing. It is easily found that
\begin{equation}
J_{CP} = \mbox{Im} V_{12}V_{23}V_{22}^{\ast}V_{13}^{\ast} \simeq \frac{1}{6}\lambda (1-\lambda^2/2-\lambda^3)\sin \delta \simeq 0.035 \sin\delta \simeq 0.035
\end{equation}

We now turn to discuss the neutrino masses. The three right-handed heavy Majorana neutrino masses are obtained
by diagonalizing the mass matrix $M_N$ with $V^{\dagger}_{\nu}
M_N V_{\nu} = diag.(m_{N_1}, \ m_{N_2},\ m_{N_3})$
\begin{eqnarray}
m_{N_1} & = & \xi^{\nu} \left(V_0^{\nu} -V_1^{\nu} + V_2^{\nu}\right) \equiv M_0, \nonumber \\
m_{N_2} & = & \xi^{\nu} \left(V_0^{\nu} + V_1^{\nu} + 2V_2^{\nu}\right) \equiv  M_0(1+\Delta_1) , \\
m_{N_3} & = & \xi^{\nu} \left(V_0^{\nu} + V_1^{\nu} - V_2^{\nu}\right) \equiv  M_0(1+\Delta_2), \nonumber
\end{eqnarray}
with
\begin{eqnarray}
\Delta_1 = \frac{\Delta_+ }{ 1 -\Delta_- },\quad
\Delta_2 =  \frac{2\Delta_-}{ 1 -\Delta_- } ; \qquad \Delta_+ \equiv \frac{ V_1^{\nu} + V_2^{\nu}}{V_0^{\nu}},\quad \Delta_- \equiv \frac{V_1^{\nu} - V_2^{\nu}}{V_0^{\nu}}
\end{eqnarray}
Thus the masses of three left-handed Majorana neutrinos are given
in the physics basis as follows
\begin{eqnarray}
& & m_{\nu_1} = m_{0} , \quad  m_{\nu_2} = m_0 \left( 1+\bar{\Delta}_1 \right),\quad  m_{\nu_3} = m_{0} \left( 1 + \bar{\Delta}_2 \right)
\end{eqnarray}
with
\begin{eqnarray}
& & m_{0} \equiv \frac{(m_{\nu}^D)^2}{M_0} =\frac{(y_L^{\nu})^2}{\xi^{\nu}} \left( \frac{v^2}{V_0^{\nu}} \right)\frac{1}{ 1 - \Delta_-  },\quad \bar{\Delta}_1 = \frac{\Delta_+}{1+\Delta_+ - \Delta_-},\quad \bar{\Delta}_2 = \frac{\Delta_-}{1+ \Delta_-}
\end{eqnarray}
Their mass-squared differences are given by
\begin{eqnarray}
\Delta m_{21}^{2}=m_{\nu_2}^2 - m_{\nu_1}^2  = 2\bar{\Delta}_1 (1 + \bar{\Delta}_1/2) m_0^2, \quad \Delta m_{31}^{2}=m_{\nu_3}^2 - m_{\nu_1}^2  = 2\bar{\Delta}_2 (1 + \bar{\Delta}_2/2) m_0^2,
\end{eqnarray}
The small ratio $|\Delta m_{21}^{2}/\Delta m_{31}^{2}| \sim \lambda^2 $ from the experimental data implies that
\begin{eqnarray}
|\bar{\Delta}_1/\bar{\Delta}_2| \ll 1, \quad |\Delta_+/\Delta_-| \ll 1, \qquad |V_1^{\nu}| \simeq -|V_2^{\nu}|
\end{eqnarray}

With the vacuum hierarchy Eq.(\ref{U1}) due to the approximate global U(1) family symmetries, and the same ansatz that the smallness of the ratios $ V_1^{\nu}/V_0^{\nu}$ and $ V_2^{\nu}/V_0^{\nu}$ is also characterized by the Wolfeinstein parameter $\lambda \simeq 0.22$, we then have
\begin{eqnarray}
|V_1^{\nu}/V_0^{\nu}|\sim \lambda, \quad |V_2^{\nu}/V_0^{\nu}| \sim \lambda,\qquad |\Delta_-| \leq 2\lambda\, .
\end{eqnarray}
To provide a prediction for neutrino masses, we shall discuss two cases: one is for the normal mass hierarchy with $\Delta_- > 0$, i.e., $\bar{\Delta}_2 > 0$, and the other is for the inverse mass hierarchy with $\Delta_- < 0$, i.e., $\bar{\Delta}_2 < 0$. By inputting the following values for the two cases
\begin{eqnarray}
\bar{\Delta}_2 =  \frac{\Delta_-}{1+ \Delta_-} \simeq \frac{4\lambda}{1+2\lambda} \simeq 0.61,\quad
\bar{\Delta}_2 =  \frac{\Delta_-}{1+ \Delta_-} \simeq -\frac{4\lambda}{1-2\lambda} \simeq -1.57
\end{eqnarray}
and using the experimentally well measured mass-squared differences  $\Delta m_{21}^{2} \simeq 7.6\times 10^{-5}$ eV$^2$ and $|\Delta m_{31}^{2}| \simeq 2.3\times 10^{-3}$ eV$^2$, we arrive at the following predictions
\begin{eqnarray}
m_{\nu_1} \simeq 3.80\times 10^{-2}\, \mbox{eV}, \quad m_{\nu_2} \simeq 3.90\times 10^{-2} \, \mbox{eV}, \quad m_{\nu_3}   \simeq 6.12 \times 10^{-2}\, \mbox{eV},
\end{eqnarray}
for the normal neutrino mass hierarchy with $\bar{\Delta}_2 \simeq 0.61$, and
\begin{eqnarray}
m_{\nu_1} \simeq 5.837 \times 10^{-2}\, \mbox{eV}, \quad m_{\nu_2}  \simeq 5.902 \times 10^{-2}\, \mbox{eV}, \quad m_{\nu_3}  \simeq 3.327\times 10^{-2}\, \mbox{eV},
\end{eqnarray}
for the inverse neutrino mass hierarchy with $\bar{\Delta}_2 \simeq -1.57$. Note that the initial mass for $m_{\nu_3}$ is negative and we have transformed it into a positive one by making a phase redefinition for the Majorana  neutrino $ \nu_3 \to i\nu_3$. It is obvious that the total neutrino mass in both cases is around
\begin{equation}
\sum m_{\nu_i} \sim 0.15 eV
\end{equation}
which is testable in future precision astrophysics and cosmology\cite{Wong:2011ip, Carbone:2011by}.

As the left-handed Majorana neutrino masses are determined by the ratio of the Dirac-type neutrino masses to the right-handed heavy Majorana neutrino masses, thus if the Dirac-type neutrino mass is at  the electroweak scale $m^D_N \sim 100$ GeV, the heavy Majorana neutrino masses should be at the order of $M_N \sim 10^{14}$ GeV. While when the Dirac-type neutrino mass is assumed to be around the electron mass $m^D_N \sim 1$ MeV, then the heavy Majorana neutrino masses could be as low as the order of $M_N \sim 10$ TeV. The heavy Majorana neutrino masses are given by the SU$_F$(3) symmetry breaking scale, thus the  SU$_F$(3) symmetry breaking scale could have a large range $10^4\sim 10^14$ GeV. On the other hand,  the  SU$_F$(3) symmetry breaking scale is directly constrained from the lepton flavor changing processes and also from the flavor changing neutral current in the quark sector. The present data on the lepton flavor changing process $\mu \to 3e$ with $Br(\mu \to 3e) < 1.0 \times 10^{-12}$ will lead the SU$_F$(3) symmetry breaking scale to be above 200 TeV. From the neutral meson mixing $K^0-\bar{K}^0$ and $B^0-\bar{B}^0$, it was shown for the case of SO(3) gauge model that the gauge symmetry breaking scale should be over 100 TeV\cite{BS}, a similar constraint is applicable to the SU$_F$(3) gauge model. Thus the  SU$_F$(3) symmetry breaking scale will set a new symmetry breaking scale above $10^3 v$ with $v=246$ GeV the electroweak scale.

In conclusion, we have provided a simple gauge model based on the SU$_F$(3) gauge family symmetry to understand the lepton-flavor mixing and masses. It has been shown that in our present model an exact tri-bimaximal mixing arises from diagonalizing the Majorana neutrino mass matrix, which is a natural consequence of a residual $Z_2$ symmetry for the  SU$_F$(3) vacuum structure in the neutrino sector, the deviation to the tri-bimaximal mixing is found to be attributed to the small mixing in the charged-lepton sector, its smallness is protected by the mechanism of approximate global U(1) family symmetries to the vacuum structure of spontaneous symmetry breaking. It is of interest to observe that with a simple ansatz that all the smallness due to the approximate global U(1) family symmetries is characterized by a single Wolfenstein parameter $\lambda \simeq 0.22$ and the spontaneous CP violation is maximal, a consistent prediction for the lepton-flavor mixing can reasonably resulted once the charged-lepton mixing matrix is taken to have a similar hierarchy structure as the CKM quark mixing matrix. In particular, the resulting $\theta_{13}$ agrees well with the most recent measurement by the Daya Bay reactor neutrino experiment\cite{An:2012eh}. Remarkably, the predicting Jarlskog-invariant for CP violation is big enough to be measured in the future experiment. The small neutrino masses can well be understood via the standard seesaw mechanism and the smallness of their  mass-squared differences is ascribed to the largely degenerate neutrino masses, which is protected again by the mechanism of approximate global U(1) family symmetries. As the neutrino masses are largely degenerate in this model, their total mass $\sum m_{\nu_i} \sim 0.15$ eV as the hot dark matter component is much larger than the minimal limit $\sum m_{\nu}\sim 0.05$ eV in models with a hierarchic neutrino mass structure, thus the present model may be tested by the future higher precision experiments in astrophysics and cosmology.

\acknowledgments

\label{ACK}

This work was supported in part by the
National Science Foundation of China (NSFC) under Grant \#No.
10821504, 10975170, and by the National Basic Research Program of China (973 Program) under Grants No. 2010CB833000, as well as the key Project of Chinese Academy of Sciences (CAS).

\end{document}